\documentclass[preprint,aps,prb,onecolumn,superscriptaddress,floatfix,longbibliography,11pt]{revtex4-2}

\usepackage{amsmath,amssymb} 
\usepackage{bm} 
\usepackage{graphicx} 
\usepackage{textcomp} 
\usepackage[nolist,nohyperlinks]{acronym} 
\usepackage{float}
\usepackage{upgreek}
\usepackage{mdframed}
\usepackage{xcolor}
\usepackage{setspace}
\usepackage{siunitx} 
\usepackage{xr} 
\usepackage{enumitem} 
\usepackage{hyperref} 
\hypersetup{
    colorlinks=true,
    urlcolor= blue,
    citecolor=blue,
    linkcolor= blue,
}

\newcommand{\mytitle}{Lasing on hybridized soliton frequency combs}

\begin{document}

\title{\mytitle}

\author{Theodore P. Letsou}
\email[]{tletsou@g.harvard.edu}
\affiliation{Harvard John A. Paulson School of Engineering and Applied Sciences, Harvard University, Cambridge, MA 02138, USA}
\affiliation{Department of Electrical Engineering and Computer Science, Massachusetts Institute of Technology, Cambridge, MA 02142, USA}

\author{Dmitry Kazakov}
\affiliation{Harvard John A. Paulson School of Engineering and Applied Sciences, Harvard University, Cambridge, MA 02138, USA}

\author{Pawan Ratra}
\affiliation{Harvard John A. Paulson School of Engineering and Applied Sciences, Harvard University, Cambridge, MA 02138, USA}
\affiliation{Department of Electrical and Electronic Engineering, Imperial College London, Exhibition Rd, South Kensington, London SW7 2BX, United Kingdom}

\author{Lorenzo L. Columbo}
\affiliation{Dipartimento di Elettronica e Telecomunicazioni, Politecnico di Torino, 10129 Torino, Italy}

\author{Massimo Brambilla}
\affiliation{Dipartimento di Fisica Interateneo, Università e Politecnico di Bari, 70126 Bari, Italy}
\affiliation{CNR-Istituto di Fotonica e Nanotecnologie, 70126 Bari, Italy}

\author{Franco Prati}
\affiliation{Dipartimento di Scienza e Alta Tecnologia, Università dell’Insubria, 22100 Como, Italy}

\author{Cristina Rimoldi}
\affiliation{Dipartimento di Elettronica e Telecomunicazioni, Politecnico di Torino, 10129 Torino, Italy}

\author{Sandro Dal Cin}
\affiliation{Institute of Solid State Electronics, TU Wien, 1040 Vienna, Austria}

\author{Nikola Opa{\v{c}}ak}
\affiliation{Institute of Solid State Electronics, TU Wien, 1040 Vienna, Austria}

\author{Henry O. Everitt}
\affiliation{DEVCOM Army Research Laboratory South, Houston, TX 77005, USA}
\affiliation{Department of Physics, Duke University, Durham, NC 27708, USA}

\author{Marco Piccardo}
\affiliation{Harvard John A. Paulson School of Engineering and Applied Sciences, Harvard University, Cambridge, MA 02138, USA}
\affiliation{Department of Physics, Instituto Superior Técnico, Universidade de Lisboa, 1049-001 Lisbon, Portugal}
\affiliation{Instituto de Engenharia de Sistemas e Computadores—Microsistemas e Nanotecnologias (INESC MN), Lisbon, Portugal}

\author{Benedikt Schwarz}
\affiliation{Harvard John A. Paulson School of Engineering and Applied Sciences, Harvard University, Cambridge, MA 02138, USA}
\affiliation{Institute of Solid State Electronics, TU Wien, 1040 Vienna, Austria}

\author{Federico Capasso}
\email[]{capasso@seas.harvard.edu}
\affiliation{Harvard John A. Paulson School of Engineering and Applied Sciences, Harvard University, Cambridge, MA 02138, USA}

\date{\today}

\begin{abstract}

\textbf{Coupling is an essential mechanism that drives complexity in natural systems, transforming single, non-interacting elements into intricate networks with rich physical properties. 
Here, we demonstrate a chip-scale coupled laser system that exhibits complex optical states impossible to achieve in an uncoupled system.
We show that a pair of coupled semiconductor ring lasers spontaneously forms a frequency comb consisting of the hybridized modes of its coupled cavity, exhibiting a large number of phase-locked tones that anticross with one another.  
Experimental coherent waveform reconstruction reveals that the hybridized frequency comb manifests itself as pairs of bright and dark picosecond-long solitons circulating simultaneously.
The dark and bright solitons exit the coupled cavity at the same time, leading to breathing bright solitons temporally overlapped with their dark soliton counterparts — a state inaccessible for a single, free-running laser.
Our results demonstrate that the rules that govern allowable states of light can be broken by simply coupling elements together, paving the way for the design of more complex networks of coupled on-chip lasers.}

\end{abstract}

\maketitle

Nonlinear interactions between elements in a system give rise to laws and phenomena that are properties of the entire body rather than just the individual elements. 
For instance, the discovery of the Fermi-Ulam-Pasta-Tsingou recurrence within a chain of nonlinear oscillators became a universal paradigm illustrating a system's return to a state near its initial condition rather than displaying ergodicity \cite{osti_4376203, Mussot2018}.
In fact, similar effects span vast areas of physics, from spontaneous pattern formation in microcavities \cite{Barland2002}, to the emergence of turbulence in fluid dynamics \cite{1883RSPT..174..935R}. 
Despite the ubiquity of these complex interactions across diverse physical systems, much remains to be explored regarding the rules and behaviors that govern these collective effects. 
Understanding how simpler systems interact provides the foundation for designing more complex, interconnected systems, such as next-generation telecommunication architectures or quantum computers \cite{Feldmann2021, Kannan2023}. 

Intriguingly, even a system of just two interacting elements can exhibit complexities that defy a reductionist ‘single-element’ description.
For example, two-component Bose-Einstein condensates (BECs) coupled to each other through the same atomic transition exhibit stable formations not accessible when the two are uncoupled \cite{Becker2008}. 
Pairs of bright and dark solitons — discrete localized structures — in the macroscopic density distributions of the coupled BECs can stabilize one another over second timescales, which would otherwise dissipate in the absence of coupling.
However, BECs are less than optimal systems to study complexity.  
They require large-scale laboratory setups to both generate and measure, limiting their versatility and flexibility.

On the other hand, semiconductor lasers are ideal platforms to study coherent nonlinear interactions in coupled networks, owing to their high output powers, diversity of available wavelengths, and most importantly, compact form factor — all while operating at room temperature.  
This allows them to be easily coupled together on the same die, making them excellent test beds for measuring complex nonlinear phenomena with a high degree of fidelity through electrical bias.
In this work, we identify a new state of light in a pair of semiconductor ring lasers with mutually coupled cavities.
Coupling two lasers together causes their cold cavity resonances to split — or hybridize — akin to atomic orbitals in molecules \cite{purcell1980introduction}.
Our coupled system spontaneously emits a frequency comb lasing on the hybridized modes of the coupled cavity under direct current (DC) injection of both lasers. 
The hybridized mode beat note frequency continuously tunes with heater current, never crossing zero, proving operation on the anticrossed resonances. At the same time, the intermode beat note of the frequency comb maintains a kHz-level linewidth, serving as proof of the mutual phase coherence of the comb lines.
Waveform reconstructions reveal that the hybridized frequency comb corresponds to a combination of bright and dark picosecond solitons in the time domain, simultaneously circulating in the coupled cavity with synchronized group velocities.
Our results show that coupling lifts the constraints of a single resonator optical system, permitting laser waveforms that would otherwise destabilize in uncoupled resonators, establishing an optical analog to coupled BECs.

The basis of our study consists of semiconductor ring lasers (laser cavities with periodic boundary conditions) with ultrashort ($\sim$ 1 ps) gain recovery times \cite{Piccardo2021_laserReview} — quantum cascade lasers (QCLs) \cite{Opaak2021, Meng2020, kazakov2021defect, Meng2021, Piccardo2020_2, Yao2012} — whose intracavity dynamics, $E(z, t)$, are well-captured by the complex Ginzburg-Landau equation (CGLE) \cite{Opaak2024, Aranson2002},
\begin{align}
\partial_tE = E  + (1 + ic_{\text{D}})\partial^2_zE - (1 + ic_{\text{NL}})|E|^2E.
\label{eq1}
\end{align}
The CGLE is a reduced laser model with a parameter space spanned by only two variables, $c_{\text{D}}$ and $c_{\text{NL}}$, which for a QCL, can be written in terms of its linewidth enhancement factor $\alpha$, group velocity dispersion $k^{''}$, and Kerr nonlinearity $\beta$~\cite{Piccardo2020_2}. 
The CGLE supports only a limited number of stable solutions for ring QCLs, such as single-mode solutions, multimode homoclons, and Nozaki-Bekki (NB) dark solitons 
 \cite{Opaak2024, Meng2021, Efremidis2000, Nozaki1984}.
Some states of light, however, are impossible to achieve in a free-running ring QCL.
One such state is a bright, background-free pulse of light, or bright soliton.
A background-free bright pulse of light must coexist with the trivial $E = 0$ solution to remain stable inside the cavity, but the presence of linear gain uniformly distributed along the ring causes noise spikes on top of the zero-field background — inevitably present in a realistic laser cavity — to amplify and destabilize the pulse \cite{Malomed1996}.

\begin{figure*}[!t]
	\includegraphics[width = 1\columnwidth]{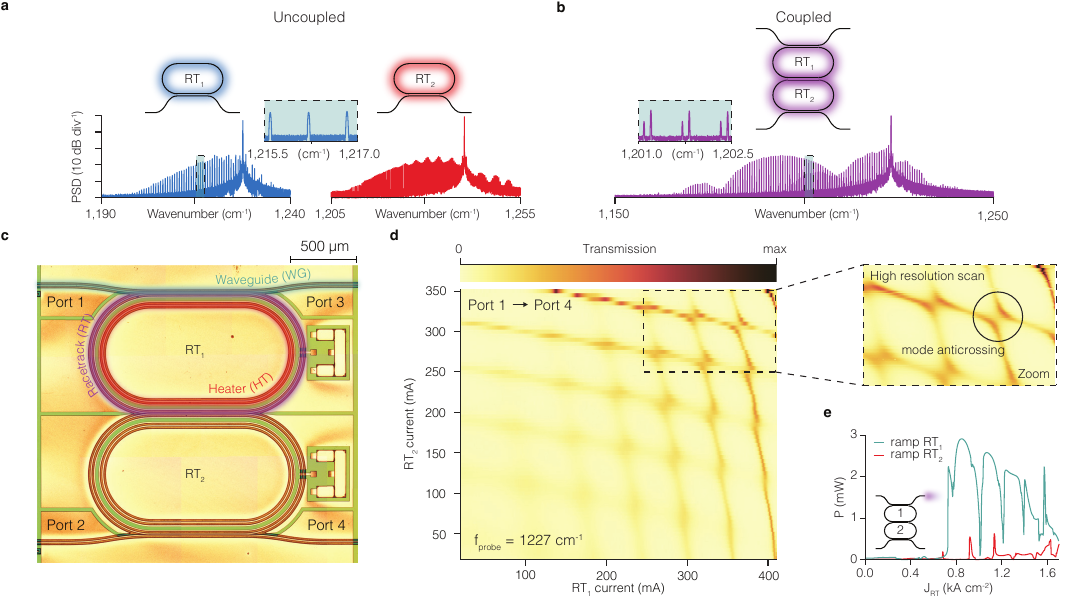}
	\caption{ \textbf{Resonant behavior of coupled racetrack quantum cascade lasers (QCLs).} \textbf{a} Uncoupled racetrack QCLs produce Nozaki-Bekki (NB) solitons, as predicted by the complex Ginzburg-Landau equation (CGLE).  \textbf{b} Two coupled racetrack QCLs emit two soliton spectra — one with a strong pump line and the other without.  The zoomed portion of the spectrum shows that the coupled racetracks lase on the hybridized resonances of the coupled cavity.  \textbf{c} shows a microscope image of the coupled racetrack QCLs used in this work, denoted as RT$_1$ and RT$_2$.  The waveguide (WG), racetrack (RT), and heater (HT) of RT$_1$ are false colored blue, purple, and red, respectively.  The four cleaved waveguide facets serve as ports to either out-couple laser light generated in RTs or to probe the system with an external light source.  \textbf{d}, The coupled laser system is probed below its lasing threshold with a tunable, single-frequency QCL injected into port 1.  The transmission of the probe laser is measured at the exit of port 4 while the bias of both RTs is swept from 20 mA to 410 mA and 350 mA for $\text{RT}_1$ and $\text{RT}_2$ respectively.  The probe laser is set to 1,227 $\text{cm}^{-1}$ — around the peak gain response of the QCL gain material. A high-resolution scan for high RT biases reveals anticrossings in the resonant structure of the coupled RTs. \textbf{e}, DC biasing both RTs above their thresholds yields mW of power at room temperature (the WGs are biased at 200 mA).}
	\label{fig1}
\end{figure*}

A pulse of light may become stable if its background field is not equal to zero \cite{Efremidis2000}.
For example, NB solitons, which are localized field excitations mixed with non-zero backgrounds, are often found to spontaneously occur in QCLs with periodic boundary conditions, such as racetrack QCLs.
In the spectral domain, they consist of many in-phase optical lines, which together form a pulse of light combined with a strong, out-of-phase, continuous-wave (CW) background, resulting in a dark pulse in the temporal domain.
This background field is often orders of magnitude stronger than the pulse, as shown by the two experimentally measured NB soliton spectra taken from two different racetrack QCLs in Fig.~\ref{fig1}\textbf{a}.
Aside from the states of complete synchronization and independent lasing, coupling two racetrack QCLs permits resonantly-mixed states not present in uncoupled racetrack QCLs.
Fig.~\ref{fig1}\textbf{b} shows an example multimode state in a coupled racetrack QCL consisting of two spectral lobes, both containing many optical lines, but only one lobe contains a CW background.
Zooming in on the optical spectrum shows that the two spectral lobes are detuned in frequency with respect to one another.
In what follows, we show the background-free lobe in this resonantly-mixed multimode spectrum corresponds to a bright soliton, while the other lobe corresponds to a dark NB soliton.

The spectrum shown in Fig.~\ref{fig1}\textbf{b} is emitted by a device experimentally realized using standard semiconductor laser fabrication processes.
We fabricate two racetrack QCLs (grown via MBE on an InP substrate) coupled along their straight sections using a dry-etch process attainable with photolithography, as shown by the microscope image in Fig.~\ref{fig1}\textbf{c} \cite{Kacmoli2023}.
Both lasers consist of a racetrack (RT, purple), a waveguide (WG, blue), and an integrated heater (HT, red), all of which can be independently addressed without electrical crosstalk under DC bias.  
Each section of the chip serves a specific purpose; the RT is used to generate laser light when biased above its threshold, the WG outcouples the laser light — yielding milliwatts of CW optical power at room temperature from one or more of its four output ports (Fig.~\ref{fig1}\textbf{e}), and biasing the HT tunes the laser resonances via a thermal refractive index change. 
The 1 $\upmu$m wide gap between the two lasers results in a measured coupling ratio of $\sim -1.25$ dB \cite{Kazakov2024}.

Coupling two optical resonators together has profound effects on the natural oscillation frequencies of the system as a whole. To experimentally demonstrate this effect, we first probe the below threshold, cold cavity modes of the coupled RTs using a single-frequency laser injected into port 1 and detected at port 4.  
The transmission of the probe laser is measured as the bias of both RTs (denoted as $\text{RT}_1$ and $\text{RT}_2$) is swept from well below their individual lasing thresholds ($\sim0.05J_{\text{th}}$) to just below their lasing thresholds ($\sim0.95J_{\text{th}}$).  
The experimental transmission map is shown in Fig.~\ref{fig1}\textbf{d}, where the area of dark brown represents high transmission through the coupled cavities.
Resistive heating sweeps the laser refractive index and therefore the frequencies of the cavity resonances.
The tilting in the heat map is caused by thermal crosstalk between the two lasers.
A finer current sweep demonstrates that the resonances of the RTs anticross with one another — a hallmark feature of mutually coupled resonators, which is absent in one-way coupled systems.
Increasing the bias of the RTs also decreases their loss through carrier population inversion, resulting in sharper, more pronounced resonance splittings, as shown from wavelength sweeps over the same resonance measured at four increasing bias settings.

\begin{figure*}[t!]
	\centering
	\includegraphics[width = 0.5\columnwidth]{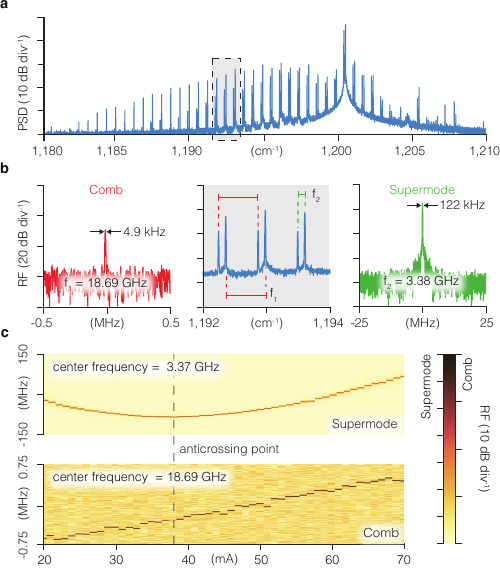}
	\caption{ \textbf{Hybridization of the frequency comb} \textbf{a}, The laser emits a coherent frequency comb over its hybridized supermodes when both RTs are biased sufficiently above their individual laser thresholds.  Splitting of the optical lines occurs due to the coupling between the RTs, which is seen in the zoomed portion of the optical spectrum.  \textbf{b}, The laser's coherence is identified by the presence of two \ac{rf} tones: the first is the `comb' beat note produced by the interaction between neighboring hybridized comb modes (FWHM = 4.9 kHz), and the second is the `supermode' beat note produced by the interaction between the split supermodes (FWHM = 122 kHz).  The beat notes have center frequencies $f_1 = 18.69$ GHz and $f_2 = 3.38$ GHz, respectively. \textbf{c}, Both \ac{rf} tones are monitored as the laser is thermally tuned using the integrated heaters.  Both beat notes are present when the system is brought to its anticrossing point.  This indicates that comb operation is maintained while the supermode spacing is tuned.
 }
	\label{fig2}
\end{figure*}

Further electrical pumping of $\text{RT}_1$ and $\text{RT}_2$ past their laser thresholds ($\sim1.5J_{\text{th}}$, and $\sim1.6J_{\text{th}}$, respectively) causes them to undergo a coherent single-mode instability, producing a frequency comb unlike ones seen in uncoupled racetrack QCLs \cite{Piccardo2020_2, Meng2020, Meng2021, Opaak2024}.  
The laser emits a frequency comb over the hybridized resonances of the coupled cavity (Fig.~\ref{fig2}\textbf{a}).  
Evidence of coherence is provided by two kHz-wide RF tones: first, the ‘comb’ beat note, produced by the mixing of adjacent comb lines at the laser repetition rate, establishes intermodal coherence (FWHM = 4.9 kHz, RBW = 5.1 kHz).  
Second, the ‘supermode’ beat note, produced by the mixing of neighboring hybridized laser lines, establishes the stability of the hybridized state (FWHM = 122 kHz, RBW = 100 kHz).  
The frequency of the ‘supermode’ beat note also measures the precise spacing between the hybridized lines.
Both beat notes (shown in Fig.~\ref{fig2}\textbf{b}) — extracted from the laser's top metal contact — are monitored while the laser resonances are thermally shifted by sweeping the bias of $\text{HT}_{2}$.
Fig.~\ref{fig2}\textbf{c} confirms comb coherence is maintained throughout the sweep range, showing the comb beat note continuously thermally tuned over $>1$ MHz to the blue.
The supermode beat note, however, does not follow the same thermal tuning pattern.
Instead, the supermode beat note shifts first towards lower frequencies, approaching $\sim$ 3.3 GHz, then shifts back to higher frequencies without ever reaching zero frequency.
This measurement demonstrates the hybridized modes of the frequency comb anticross while maintaining a high level of intermodal coherence.

\begin{figure*}[t!]
	\centering
	\includegraphics[width = 1\columnwidth]{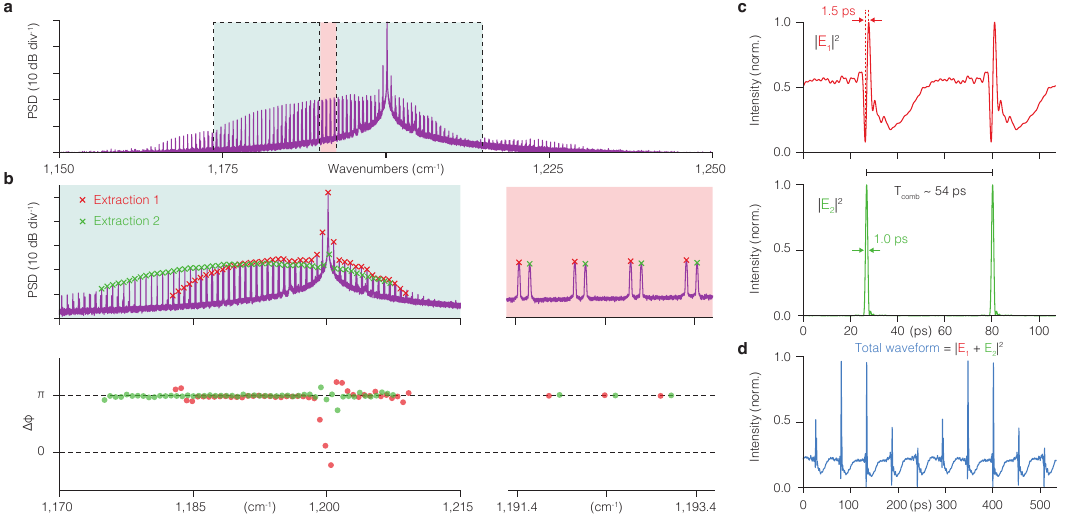}
	\caption{ \textbf{Waveform analysis of the hybridized frequency comb.} \textbf{a}, The optical spectrum of the hybridized comb state contains two groups of modes, each spaced by the same intermode frequency.  The directionality of the laser state is deduced by measuring light output at all four ports of the coupled RTs, noting strong outputs only at ports 3 and 4. Panel \textbf{b} shows two zoomed portions of the optical spectrum with two mode groups marked with green and red crosses.  Each pair of modes produces a beating and their difference frequency, and the phase of the beating — which is the phase difference between subsequent pairs of comb teeth, $\Delta\phi$ — is measured using SWIFTS.  Both groupings of modes are completely in phase with one another, as shown from the flat $\Delta\phi$ for each extraction, with the exception of the strongest mode in the optical spectrum — the CW background.  The background is $\pi$ shifted with respect to all the other modes, and distorts the phase of its neighboring modes.  \textbf{c}, Reconstruction of the waveform is achieved through cumulative summation of all the intermodal phases (along with the optical spectrum).  Reconstruction is done sequentially for each mode grouping.  The red modes correspond to a 1.5 ps dark pulse of light, while the green modes correspond to a 1.0 ps bright pulse of light.  \textbf{d}, The total waveform of the hybridized comb is the complex addition of the two previously calculated waveforms, leading to a time-domain profile where the bright and dark pulses overlap with one another.  The supermode spacing — the spacing between the hybridized modes — defines the periodicity of the “breathing" of the bright pulse atop the CW background, indicated by the dashed curve.}
	\label{fig3}
\end{figure*}

The intricate balance between the complete locking of each RT's laser lines with the repulsive action of strongly-coupled resonators constitutes a new form of laser mode-locking.
As a result, its temporal waveform differs greatly from other mode-locked lasers and QCL frequency combs.
We use the SWIFTS waveform reconstruction technique to analyze the hybridized frequency comb in the time domain \cite{Burghoff2015}.
SWIFTS is well-suited for mode-locked lasers emitting in the mid-infrared as it does not require an optical nonlinearity for waveform reconstruction like FROG, IAC, or SPIDER \cite{Stibenz2006}; rather, it relies on quadrature detection of the laser beat note on a fast photodetector, allowing time-domain analysis of both pulsed and non-pulsed waveforms \cite{han2020sensitivity}.
We measure the waveform of the hybridized-locked comb state in Fig.~\ref{fig3}\textbf{a}, whose unidirectional emission in each RT causes light to be measured only from ports 3 and 4. 
Much like the states shown in Fig.~\ref{fig1}\textbf{b} and Fig.~\ref{fig2}\textbf{a}, its optical spectrum consists of two groups of intertwined modes (marked green and red in the zoomed area of Fig.~\ref{fig3}\textbf{b}) split by the mode hybridization.
The mode groupings share the same repetition rate; therefore, all intermode beat note phases — i.e., the phase differences between pairs of adjacent comb lines — are measured in a single SWIFTS acquisition.
Waveform reconstruction is performed sequentially on each set of modes.
Waveform reconstruction on the first family of modes in red yields a $\sim$ 1.5 ps dark NB soliton, while the second family of modes shown in green corresponds to a $\sim$ 1.0 ps bright soliton with the same round-trip time as the NB soliton (Fig.~\ref{fig3}\textbf{c}).
Unlike the NB soliton, however, the bright soliton is not a stable solution to the cubic CGLE and can only exist when coupled to the NB soliton.
The total waveform is the complex addition of the two reconstructions, which results in a mixture of dark-bright solitons breathing atop the CW background, as seen in Fig.~\ref{fig3}\textbf{d}.

Coupling two ring lasers together marks the simplest possible deviation from a single-resonator model.
As such, single-field theories are replaced with their coupled counterparts, which contain more stable solutions not achievable in isolated geometries \cite{busch2001dark}.
This insight, combined with the self-starting frequency comb capabilities of the quantum cascade architecture, unlocks new states of light, which not only serve as optical analogs for complexity but stand alone as useful outputs.
For example, the presented system could serve as an excellent platform for miniaturizing non-Hermitian photonic devices, such as exceptional-point-enhanced ring laser gyroscopes \cite{Hokmabadi2019} — or could even act as a compact dual-comb spectrometer, where the comb repetition rate in each racetrack can be detuned and locked to an external reference synthesizer \cite{Picqu2019}.

In addition, coupling lasers has shown promise in generating topologically protected edge pathways for light to travel \cite{mittal2021topological, Choi2021, Bandres2018}.
Combining such effects with frequency combs opens the door to studying multimode topological photonics \cite{Flower2024}.
As more ring lasers are connected together, we expect them to form a continuum of optical states, with photonic bands separated by photonic bandgaps — much like a photonic crystal.
We foresee a plethora of future active devices embracing coupling as a new design parameter: many rings coupled together, arrays of rings, rings connected with passive waveguides — all integrated on the same chip, with each arrangement of components governed by its own set of coupled equations, possibly superseding the current understanding of allowable states in semiconductor lasers.

This material is based on work supported by the National Science Foundation under Grant No. ECCS-2221715. T. P. Letsou thanks the Department of Defense (DoD) through the National Defense Science and Engineering Graduate (NDSEG) Fellowship Program. N. Opa$\mathrm{\check{c}}$ak and B. Schwarz received funding from the European Research Council (ERC) under the European Union’s Horizon 2020 research and innovation program (Grant agreement No. 853014).
The authors gratefully acknowledge the Center for Micro- and Nanostructures (ZMNS) of TU Wien for providing the cleanroom facilities.

\newpage

\bibliography{coupledRings}

\begin{thebibliography}{32}%
\makeatletter
\providecommand \@ifxundefined [1]{%
 \@ifx{#1\undefined}
}%
\providecommand \@ifnum [1]{%
 \ifnum #1\expandafter \@firstoftwo
 \else \expandafter \@secondoftwo
 \fi
}%
\providecommand \@ifx [1]{%
 \ifx #1\expandafter \@firstoftwo
 \else \expandafter \@secondoftwo
 \fi
}%
\providecommand \natexlab [1]{#1}%
\providecommand \enquote  [1]{``#1''}%
\providecommand \bibnamefont  [1]{#1}%
\providecommand \bibfnamefont [1]{#1}%
\providecommand \citenamefont [1]{#1}%
\providecommand \href@noop [0]{\@secondoftwo}%
\providecommand \href [0]{\begingroup \@sanitize@url \@href}%
\providecommand \@href[1]{\@@startlink{#1}\@@href}%
\providecommand \@@href[1]{\endgroup#1\@@endlink}%
\providecommand \@sanitize@url [0]{\catcode `\\12\catcode `\$12\catcode `\&12\catcode `\#12\catcode `\^12\catcode `\_12\catcode `\%12\relax}%
\providecommand \@@startlink[1]{}%
\providecommand \@@endlink[0]{}%
\providecommand \url  [0]{\begingroup\@sanitize@url \@url }%
\providecommand \@url [1]{\endgroup\@href {#1}{\urlprefix }}%
\providecommand \urlprefix  [0]{URL }%
\providecommand \Eprint [0]{\href }%
\providecommand \doibase [0]{https://doi.org/}%
\providecommand \selectlanguage [0]{\@gobble}%
\providecommand \bibinfo  [0]{\@secondoftwo}%
\providecommand \bibfield  [0]{\@secondoftwo}%
\providecommand \translation [1]{[#1]}%
\providecommand \BibitemOpen [0]{}%
\providecommand \bibitemStop [0]{}%
\providecommand \bibitemNoStop [0]{.\EOS\space}%
\providecommand \EOS [0]{\spacefactor3000\relax}%
\providecommand \BibitemShut  [1]{\csname bibitem#1\endcsname}%
\let\auto@bib@innerbib\@empty
\bibitem [{\citenamefont {Fermi}\ \emph {et~al.}(1955)\citenamefont {Fermi}, \citenamefont {Pasta}, \citenamefont {Ulam},\ and\ \citenamefont {Tsingou}}]{osti_4376203}%
  \BibitemOpen
  \bibfield  {author} {\bibinfo {author} {\bibfnamefont {E.}~\bibnamefont {Fermi}}, \bibinfo {author} {\bibfnamefont {P.}~\bibnamefont {Pasta}}, \bibinfo {author} {\bibfnamefont {S.}~\bibnamefont {Ulam}},\ and\ \bibinfo {author} {\bibfnamefont {M.}~\bibnamefont {Tsingou}},\ }\bibfield  {title} {\bibinfo {title} {Studies of the nonlinear problems}\ }\href {https://doi.org/10.2172/4376203} {10.2172/4376203} (\bibinfo {year} {1955})\BibitemShut {NoStop}%
\bibitem [{\citenamefont {Mussot}\ \emph {et~al.}(2018)\citenamefont {Mussot}, \citenamefont {Naveau}, \citenamefont {Conforti}, \citenamefont {Kudlinski}, \citenamefont {Copie}, \citenamefont {Szriftgiser},\ and\ \citenamefont {Trillo}}]{Mussot2018}%
  \BibitemOpen
  \bibfield  {author} {\bibinfo {author} {\bibfnamefont {A.}~\bibnamefont {Mussot}}, \bibinfo {author} {\bibfnamefont {C.}~\bibnamefont {Naveau}}, \bibinfo {author} {\bibfnamefont {M.}~\bibnamefont {Conforti}}, \bibinfo {author} {\bibfnamefont {A.}~\bibnamefont {Kudlinski}}, \bibinfo {author} {\bibfnamefont {F.}~\bibnamefont {Copie}}, \bibinfo {author} {\bibfnamefont {P.}~\bibnamefont {Szriftgiser}},\ and\ \bibinfo {author} {\bibfnamefont {S.}~\bibnamefont {Trillo}},\ }\bibfield  {title} {\bibinfo {title} {Fibre multi-wave mixing combs reveal the broken symmetry of fermi–pasta–ulam recurrence},\ }\href {https://doi.org/10.1038/s41566-018-0136-1} {\bibfield  {journal} {\bibinfo  {journal} {Nature Photonics}\ }\textbf {\bibinfo {volume} {12}},\ \bibinfo {pages} {303–308} (\bibinfo {year} {2018})}\BibitemShut {NoStop}%
\bibitem [{\citenamefont {Barland}\ \emph {et~al.}(2002)\citenamefont {Barland}, \citenamefont {Tredicce}, \citenamefont {Brambilla}, \citenamefont {Lugiato}, \citenamefont {Balle}, \citenamefont {Giudici}, \citenamefont {Maggipinto}, \citenamefont {Spinelli}, \citenamefont {Tissoni}, \citenamefont {Kn\"{o}dl}, \citenamefont {Miller},\ and\ \citenamefont {J\"{a}ger}}]{Barland2002}%
  \BibitemOpen
  \bibfield  {author} {\bibinfo {author} {\bibfnamefont {S.}~\bibnamefont {Barland}}, \bibinfo {author} {\bibfnamefont {J.~R.}\ \bibnamefont {Tredicce}}, \bibinfo {author} {\bibfnamefont {M.}~\bibnamefont {Brambilla}}, \bibinfo {author} {\bibfnamefont {L.~A.}\ \bibnamefont {Lugiato}}, \bibinfo {author} {\bibfnamefont {S.}~\bibnamefont {Balle}}, \bibinfo {author} {\bibfnamefont {M.}~\bibnamefont {Giudici}}, \bibinfo {author} {\bibfnamefont {T.}~\bibnamefont {Maggipinto}}, \bibinfo {author} {\bibfnamefont {L.}~\bibnamefont {Spinelli}}, \bibinfo {author} {\bibfnamefont {G.}~\bibnamefont {Tissoni}}, \bibinfo {author} {\bibfnamefont {T.}~\bibnamefont {Kn\"{o}dl}}, \bibinfo {author} {\bibfnamefont {M.}~\bibnamefont {Miller}},\ and\ \bibinfo {author} {\bibfnamefont {R.}~\bibnamefont {J\"{a}ger}},\ }\bibfield  {title} {\bibinfo {title} {Cavity solitons as pixels in semiconductor microcavities},\ }\href {https://doi.org/10.1038/nature01049} {\bibfield  {journal} {\bibinfo  {journal} {Nature}\ }\textbf {\bibinfo
  {volume} {419}},\ \bibinfo {pages} {699–702} (\bibinfo {year} {2002})}\BibitemShut {NoStop}%
\bibitem [{\citenamefont {{Reynolds}}(1883)}]{1883RSPT..174..935R}%
  \BibitemOpen
  \bibfield  {author} {\bibinfo {author} {\bibfnamefont {O.}~\bibnamefont {{Reynolds}}},\ }\bibfield  {title} {\bibinfo {title} {{An Experimental Investigation of the Circumstances Which Determine Whether the Motion of Water Shall Be Direct or Sinuous, and of the Law of Resistance in Parallel Channels}},\ }\href@noop {} {\bibfield  {journal} {\bibinfo  {journal} {Philosophical Transactions of the Royal Society of London Series I}\ }\textbf {\bibinfo {volume} {174}},\ \bibinfo {pages} {935} (\bibinfo {year} {1883})}\BibitemShut {NoStop}%
\bibitem [{\citenamefont {Feldmann}\ \emph {et~al.}(2021)\citenamefont {Feldmann}, \citenamefont {Youngblood}, \citenamefont {Karpov}, \citenamefont {Gehring}, \citenamefont {Li}, \citenamefont {Stappers}, \citenamefont {Gallo}, \citenamefont {Fu}, \citenamefont {Lukashchuk}, \citenamefont {Raja}, \citenamefont {Liu}, \citenamefont {Wright}, \citenamefont {Sebastian}, \citenamefont {Kippenberg}, \citenamefont {Pernice},\ and\ \citenamefont {Bhaskaran}}]{Feldmann2021}%
  \BibitemOpen
  \bibfield  {author} {\bibinfo {author} {\bibfnamefont {J.}~\bibnamefont {Feldmann}}, \bibinfo {author} {\bibfnamefont {N.}~\bibnamefont {Youngblood}}, \bibinfo {author} {\bibfnamefont {M.}~\bibnamefont {Karpov}}, \bibinfo {author} {\bibfnamefont {H.}~\bibnamefont {Gehring}}, \bibinfo {author} {\bibfnamefont {X.}~\bibnamefont {Li}}, \bibinfo {author} {\bibfnamefont {M.}~\bibnamefont {Stappers}}, \bibinfo {author} {\bibfnamefont {M.~L.}\ \bibnamefont {Gallo}}, \bibinfo {author} {\bibfnamefont {X.}~\bibnamefont {Fu}}, \bibinfo {author} {\bibfnamefont {A.}~\bibnamefont {Lukashchuk}}, \bibinfo {author} {\bibfnamefont {A.~S.}\ \bibnamefont {Raja}}, \bibinfo {author} {\bibfnamefont {J.}~\bibnamefont {Liu}}, \bibinfo {author} {\bibfnamefont {C.~D.}\ \bibnamefont {Wright}}, \bibinfo {author} {\bibfnamefont {A.}~\bibnamefont {Sebastian}}, \bibinfo {author} {\bibfnamefont {T.~J.}\ \bibnamefont {Kippenberg}}, \bibinfo {author} {\bibfnamefont {W.~H.~P.}\ \bibnamefont {Pernice}},\ and\ \bibinfo {author} {\bibfnamefont
  {H.}~\bibnamefont {Bhaskaran}},\ }\bibfield  {title} {\bibinfo {title} {Parallel convolutional processing using an integrated photonic tensor core},\ }\href {https://doi.org/10.1038/s41586-020-03070-1} {\bibfield  {journal} {\bibinfo  {journal} {Nature}\ }\textbf {\bibinfo {volume} {589}},\ \bibinfo {pages} {52} (\bibinfo {year} {2021})}\BibitemShut {NoStop}%
\bibitem [{\citenamefont {Kannan}\ \emph {et~al.}(2023)\citenamefont {Kannan}, \citenamefont {Almanakly}, \citenamefont {Sung}, \citenamefont {Paolo}, \citenamefont {Rower}, \citenamefont {Braum\"{u}ller}, \citenamefont {Melville}, \citenamefont {Niedzielski}, \citenamefont {Karamlou}, \citenamefont {Serniak}, \citenamefont {Veps\"{a}l\"{a}inen}, \citenamefont {Schwartz}, \citenamefont {Yoder}, \citenamefont {Winik}, \citenamefont {Wang}, \citenamefont {Orlando}, \citenamefont {Gustavsson}, \citenamefont {Grover},\ and\ \citenamefont {Oliver}}]{Kannan2023}%
  \BibitemOpen
  \bibfield  {author} {\bibinfo {author} {\bibfnamefont {B.}~\bibnamefont {Kannan}}, \bibinfo {author} {\bibfnamefont {A.}~\bibnamefont {Almanakly}}, \bibinfo {author} {\bibfnamefont {Y.}~\bibnamefont {Sung}}, \bibinfo {author} {\bibfnamefont {A.~D.}\ \bibnamefont {Paolo}}, \bibinfo {author} {\bibfnamefont {D.~A.}\ \bibnamefont {Rower}}, \bibinfo {author} {\bibfnamefont {J.}~\bibnamefont {Braum\"{u}ller}}, \bibinfo {author} {\bibfnamefont {A.}~\bibnamefont {Melville}}, \bibinfo {author} {\bibfnamefont {B.~M.}\ \bibnamefont {Niedzielski}}, \bibinfo {author} {\bibfnamefont {A.}~\bibnamefont {Karamlou}}, \bibinfo {author} {\bibfnamefont {K.}~\bibnamefont {Serniak}}, \bibinfo {author} {\bibfnamefont {A.}~\bibnamefont {Veps\"{a}l\"{a}inen}}, \bibinfo {author} {\bibfnamefont {M.~E.}\ \bibnamefont {Schwartz}}, \bibinfo {author} {\bibfnamefont {J.~L.}\ \bibnamefont {Yoder}}, \bibinfo {author} {\bibfnamefont {R.}~\bibnamefont {Winik}}, \bibinfo {author} {\bibfnamefont {J.~I.-J.}\ \bibnamefont {Wang}}, \bibinfo
  {author} {\bibfnamefont {T.~P.}\ \bibnamefont {Orlando}}, \bibinfo {author} {\bibfnamefont {S.}~\bibnamefont {Gustavsson}}, \bibinfo {author} {\bibfnamefont {J.~A.}\ \bibnamefont {Grover}},\ and\ \bibinfo {author} {\bibfnamefont {W.~D.}\ \bibnamefont {Oliver}},\ }\bibfield  {title} {\bibinfo {title} {On-demand directional microwave photon emission using waveguide quantum electrodynamics},\ }\href {https://doi.org/10.1038/s41567-022-01869-5} {\bibfield  {journal} {\bibinfo  {journal} {Nature Physics}\ }\textbf {\bibinfo {volume} {19}},\ \bibinfo {pages} {394} (\bibinfo {year} {2023})}\BibitemShut {NoStop}%
\bibitem [{\citenamefont {Becker}\ \emph {et~al.}(2008)\citenamefont {Becker}, \citenamefont {Stellmer}, \citenamefont {Soltan-Panahi}, \citenamefont {D\"{o}rscher}, \citenamefont {Baumert}, \citenamefont {Richter}, \citenamefont {Kronj\"{a}ger}, \citenamefont {Bongs},\ and\ \citenamefont {Sengstock}}]{Becker2008}%
  \BibitemOpen
  \bibfield  {author} {\bibinfo {author} {\bibfnamefont {C.}~\bibnamefont {Becker}}, \bibinfo {author} {\bibfnamefont {S.}~\bibnamefont {Stellmer}}, \bibinfo {author} {\bibfnamefont {P.}~\bibnamefont {Soltan-Panahi}}, \bibinfo {author} {\bibfnamefont {S.}~\bibnamefont {D\"{o}rscher}}, \bibinfo {author} {\bibfnamefont {M.}~\bibnamefont {Baumert}}, \bibinfo {author} {\bibfnamefont {E.-M.}\ \bibnamefont {Richter}}, \bibinfo {author} {\bibfnamefont {J.}~\bibnamefont {Kronj\"{a}ger}}, \bibinfo {author} {\bibfnamefont {K.}~\bibnamefont {Bongs}},\ and\ \bibinfo {author} {\bibfnamefont {K.}~\bibnamefont {Sengstock}},\ }\bibfield  {title} {\bibinfo {title} {Oscillations and interactions of dark and dark–bright solitons in bose–einstein condensates},\ }\href {https://doi.org/10.1038/nphys962} {\bibfield  {journal} {\bibinfo  {journal} {Nature Physics}\ }\textbf {\bibinfo {volume} {4}},\ \bibinfo {pages} {496–501} (\bibinfo {year} {2008})}\BibitemShut {NoStop}%
\bibitem [{\citenamefont {Purcell}\ and\ \citenamefont {Kotz}(1980)}]{purcell1980introduction}%
  \BibitemOpen
  \bibfield  {author} {\bibinfo {author} {\bibfnamefont {K.~F.}\ \bibnamefont {Purcell}}\ and\ \bibinfo {author} {\bibfnamefont {J.~C.}\ \bibnamefont {Kotz}},\ }\bibfield  {title} {\bibinfo {title} {An introduction to inorganic chemistry},\ }\href@noop {} {\  (\bibinfo {year} {1980})}\BibitemShut {NoStop}%
\bibitem [{\citenamefont {Piccardo}\ and\ \citenamefont {Capasso}(2021)}]{Piccardo2021_laserReview}%
  \BibitemOpen
  \bibfield  {author} {\bibinfo {author} {\bibfnamefont {M.}~\bibnamefont {Piccardo}}\ and\ \bibinfo {author} {\bibfnamefont {F.}~\bibnamefont {Capasso}},\ }\bibfield  {title} {\bibinfo {title} {Laser frequency combs with fast gain recovery: Physics and applications},\ }\href {https://doi.org/10.1002/lpor.202100403} {\bibfield  {journal} {\bibinfo  {journal} {Laser and Photonics Reviews}\ }\textbf {\bibinfo {volume} {16}} (\bibinfo {year} {2021})}\BibitemShut {NoStop}%
\bibitem [{\citenamefont {Opa{\v{c}}ak}\ \emph {et~al.}(2021)\citenamefont {Opa{\v{c}}ak}, \citenamefont {Pilat}, \citenamefont {Kazakov}, \citenamefont {Cin}, \citenamefont {Ramer}, \citenamefont {Lendl}, \citenamefont {Capasso},\ and\ \citenamefont {Schwarz}}]{Opaak2021}%
  \BibitemOpen
  \bibfield  {author} {\bibinfo {author} {\bibfnamefont {N.}~\bibnamefont {Opa{\v{c}}ak}}, \bibinfo {author} {\bibfnamefont {F.}~\bibnamefont {Pilat}}, \bibinfo {author} {\bibfnamefont {D.}~\bibnamefont {Kazakov}}, \bibinfo {author} {\bibfnamefont {S.~D.}\ \bibnamefont {Cin}}, \bibinfo {author} {\bibfnamefont {G.}~\bibnamefont {Ramer}}, \bibinfo {author} {\bibfnamefont {B.}~\bibnamefont {Lendl}}, \bibinfo {author} {\bibfnamefont {F.}~\bibnamefont {Capasso}},\ and\ \bibinfo {author} {\bibfnamefont {B.}~\bibnamefont {Schwarz}},\ }\bibfield  {title} {\bibinfo {title} {Spectrally resolved linewidth enhancement factor of a semiconductor frequency comb},\ }\href {https://doi.org/10.1364/optica.428096} {\bibfield  {journal} {\bibinfo  {journal} {Optica}\ }\textbf {\bibinfo {volume} {8}},\ \bibinfo {pages} {1227} (\bibinfo {year} {2021})}\BibitemShut {NoStop}%
\bibitem [{\citenamefont {Meng}\ \emph {et~al.}(2020)\citenamefont {Meng}, \citenamefont {Singleton}, \citenamefont {Shahmohammadi}, \citenamefont {Kapsalidis}, \citenamefont {Wang}, \citenamefont {Beck},\ and\ \citenamefont {Faist}}]{Meng2020}%
  \BibitemOpen
  \bibfield  {author} {\bibinfo {author} {\bibfnamefont {B.}~\bibnamefont {Meng}}, \bibinfo {author} {\bibfnamefont {M.}~\bibnamefont {Singleton}}, \bibinfo {author} {\bibfnamefont {M.}~\bibnamefont {Shahmohammadi}}, \bibinfo {author} {\bibfnamefont {F.}~\bibnamefont {Kapsalidis}}, \bibinfo {author} {\bibfnamefont {R.}~\bibnamefont {Wang}}, \bibinfo {author} {\bibfnamefont {M.}~\bibnamefont {Beck}},\ and\ \bibinfo {author} {\bibfnamefont {J.}~\bibnamefont {Faist}},\ }\bibfield  {title} {\bibinfo {title} {Mid-infrared frequency comb from a ring quantum cascade laser},\ }\href {https://doi.org/10.1364/optica.377755} {\bibfield  {journal} {\bibinfo  {journal} {Optica}\ }\textbf {\bibinfo {volume} {7}},\ \bibinfo {pages} {162} (\bibinfo {year} {2020})}\BibitemShut {NoStop}%
\bibitem [{\citenamefont {Kazakov}\ \emph {et~al.}(2021)\citenamefont {Kazakov}, \citenamefont {Opačak}, \citenamefont {Beiser}, \citenamefont {Belyanin}, \citenamefont {Schwarz}, \citenamefont {Piccardo},\ and\ \citenamefont {Capasso}}]{kazakov2021defect}%
  \BibitemOpen
  \bibfield  {author} {\bibinfo {author} {\bibfnamefont {D.}~\bibnamefont {Kazakov}}, \bibinfo {author} {\bibfnamefont {N.}~\bibnamefont {Opačak}}, \bibinfo {author} {\bibfnamefont {M.}~\bibnamefont {Beiser}}, \bibinfo {author} {\bibfnamefont {A.}~\bibnamefont {Belyanin}}, \bibinfo {author} {\bibfnamefont {B.}~\bibnamefont {Schwarz}}, \bibinfo {author} {\bibfnamefont {M.}~\bibnamefont {Piccardo}},\ and\ \bibinfo {author} {\bibfnamefont {F.}~\bibnamefont {Capasso}},\ }\bibfield  {title} {\bibinfo {title} {Defect-engineered ring laser harmonic frequency combs},\ }\href {https://doi.org/10.1364/optica.430896} {\bibfield  {journal} {\bibinfo  {journal} {Optica}\ }\textbf {\bibinfo {volume} {8}},\ \bibinfo {pages} {1277} (\bibinfo {year} {2021})}\BibitemShut {NoStop}%
\bibitem [{\citenamefont {Meng}\ \emph {et~al.}(2021)\citenamefont {Meng}, \citenamefont {Singleton}, \citenamefont {Hillbrand}, \citenamefont {Franckié}, \citenamefont {Beck},\ and\ \citenamefont {Faist}}]{Meng2021}%
  \BibitemOpen
  \bibfield  {author} {\bibinfo {author} {\bibfnamefont {B.}~\bibnamefont {Meng}}, \bibinfo {author} {\bibfnamefont {M.}~\bibnamefont {Singleton}}, \bibinfo {author} {\bibfnamefont {J.}~\bibnamefont {Hillbrand}}, \bibinfo {author} {\bibfnamefont {M.}~\bibnamefont {Franckié}}, \bibinfo {author} {\bibfnamefont {M.}~\bibnamefont {Beck}},\ and\ \bibinfo {author} {\bibfnamefont {J.}~\bibnamefont {Faist}},\ }\bibfield  {title} {\bibinfo {title} {Dissipative kerr solitons in semiconductor ring lasers},\ }\href {https://doi.org/10.1038/s41566-021-00927-3} {\bibfield  {journal} {\bibinfo  {journal} {Nature Photonics}\ }\textbf {\bibinfo {volume} {16}},\ \bibinfo {pages} {142–147} (\bibinfo {year} {2021})}\BibitemShut {NoStop}%
\bibitem [{\citenamefont {Piccardo}\ \emph {et~al.}(2020)\citenamefont {Piccardo}, \citenamefont {Schwarz}, \citenamefont {Kazakov}, \citenamefont {Beiser}, \citenamefont {Opa{\v{c}}ak}, \citenamefont {Wang}, \citenamefont {Jha}, \citenamefont {Hillbrand}, \citenamefont {Tamagnone}, \citenamefont {Chen}, \citenamefont {Zhu}, \citenamefont {Columbo}, \citenamefont {Belyanin},\ and\ \citenamefont {Capasso}}]{Piccardo2020_2}%
  \BibitemOpen
  \bibfield  {author} {\bibinfo {author} {\bibfnamefont {M.}~\bibnamefont {Piccardo}}, \bibinfo {author} {\bibfnamefont {B.}~\bibnamefont {Schwarz}}, \bibinfo {author} {\bibfnamefont {D.}~\bibnamefont {Kazakov}}, \bibinfo {author} {\bibfnamefont {M.}~\bibnamefont {Beiser}}, \bibinfo {author} {\bibfnamefont {N.}~\bibnamefont {Opa{\v{c}}ak}}, \bibinfo {author} {\bibfnamefont {Y.}~\bibnamefont {Wang}}, \bibinfo {author} {\bibfnamefont {S.}~\bibnamefont {Jha}}, \bibinfo {author} {\bibfnamefont {J.}~\bibnamefont {Hillbrand}}, \bibinfo {author} {\bibfnamefont {M.}~\bibnamefont {Tamagnone}}, \bibinfo {author} {\bibfnamefont {W.~T.}\ \bibnamefont {Chen}}, \bibinfo {author} {\bibfnamefont {A.~Y.}\ \bibnamefont {Zhu}}, \bibinfo {author} {\bibfnamefont {L.~L.}\ \bibnamefont {Columbo}}, \bibinfo {author} {\bibfnamefont {A.}~\bibnamefont {Belyanin}},\ and\ \bibinfo {author} {\bibfnamefont {F.}~\bibnamefont {Capasso}},\ }\bibfield  {title} {\bibinfo {title} {Frequency combs induced by phase turbulence},\ }\href
  {https://doi.org/10.1038/s41586-020-2386-6} {\bibfield  {journal} {\bibinfo  {journal} {Nature}\ }\textbf {\bibinfo {volume} {582}},\ \bibinfo {pages} {360} (\bibinfo {year} {2020})}\BibitemShut {NoStop}%
\bibitem [{\citenamefont {Yao}\ \emph {et~al.}(2012)\citenamefont {Yao}, \citenamefont {Hoffman},\ and\ \citenamefont {Gmachl}}]{Yao2012}%
  \BibitemOpen
  \bibfield  {author} {\bibinfo {author} {\bibfnamefont {Y.}~\bibnamefont {Yao}}, \bibinfo {author} {\bibfnamefont {A.~J.}\ \bibnamefont {Hoffman}},\ and\ \bibinfo {author} {\bibfnamefont {C.~F.}\ \bibnamefont {Gmachl}},\ }\bibfield  {title} {\bibinfo {title} {Mid-infrared quantum cascade lasers},\ }\href {https://doi.org/10.1038/nphoton.2012.143} {\bibfield  {journal} {\bibinfo  {journal} {Nature Photonics}\ }\textbf {\bibinfo {volume} {6}},\ \bibinfo {pages} {432} (\bibinfo {year} {2012})}\BibitemShut {NoStop}%
\bibitem [{\citenamefont {Opačak}\ \emph {et~al.}(2024)\citenamefont {Opačak}, \citenamefont {Kazakov}, \citenamefont {Columbo}, \citenamefont {Beiser}, \citenamefont {Letsou}, \citenamefont {Pilat}, \citenamefont {Brambilla}, \citenamefont {Prati}, \citenamefont {Piccardo}, \citenamefont {Capasso},\ and\ \citenamefont {Schwarz}}]{Opaak2024}%
  \BibitemOpen
  \bibfield  {author} {\bibinfo {author} {\bibfnamefont {N.}~\bibnamefont {Opačak}}, \bibinfo {author} {\bibfnamefont {D.}~\bibnamefont {Kazakov}}, \bibinfo {author} {\bibfnamefont {L.~L.}\ \bibnamefont {Columbo}}, \bibinfo {author} {\bibfnamefont {M.}~\bibnamefont {Beiser}}, \bibinfo {author} {\bibfnamefont {T.~P.}\ \bibnamefont {Letsou}}, \bibinfo {author} {\bibfnamefont {F.}~\bibnamefont {Pilat}}, \bibinfo {author} {\bibfnamefont {M.}~\bibnamefont {Brambilla}}, \bibinfo {author} {\bibfnamefont {F.}~\bibnamefont {Prati}}, \bibinfo {author} {\bibfnamefont {M.}~\bibnamefont {Piccardo}}, \bibinfo {author} {\bibfnamefont {F.}~\bibnamefont {Capasso}},\ and\ \bibinfo {author} {\bibfnamefont {B.}~\bibnamefont {Schwarz}},\ }\bibfield  {title} {\bibinfo {title} {Nozaki–bekki solitons in semiconductor lasers},\ }\href {https://doi.org/10.1038/s41586-023-06915-7} {\bibfield  {journal} {\bibinfo  {journal} {Nature}\ }\textbf {\bibinfo {volume} {625}},\ \bibinfo {pages} {685–690} (\bibinfo {year}
  {2024})}\BibitemShut {NoStop}%
\bibitem [{\citenamefont {Aranson}\ and\ \citenamefont {Kramer}(2002)}]{Aranson2002}%
  \BibitemOpen
  \bibfield  {author} {\bibinfo {author} {\bibfnamefont {I.~S.}\ \bibnamefont {Aranson}}\ and\ \bibinfo {author} {\bibfnamefont {L.}~\bibnamefont {Kramer}},\ }\bibfield  {title} {\bibinfo {title} {The world of the complex ginzburg-landau equation},\ }\href {https://doi.org/10.1103/revmodphys.74.99} {\bibfield  {journal} {\bibinfo  {journal} {Reviews of Modern Physics}\ }\textbf {\bibinfo {volume} {74}},\ \bibinfo {pages} {99} (\bibinfo {year} {2002})}\BibitemShut {NoStop}%
\bibitem [{\citenamefont {Efremidis}\ \emph {et~al.}(2000)\citenamefont {Efremidis}, \citenamefont {Hizanidis}, \citenamefont {Nistazakis}, \citenamefont {Frantzeskakis},\ and\ \citenamefont {Malomed}}]{Efremidis2000}%
  \BibitemOpen
  \bibfield  {author} {\bibinfo {author} {\bibfnamefont {N.}~\bibnamefont {Efremidis}}, \bibinfo {author} {\bibfnamefont {K.}~\bibnamefont {Hizanidis}}, \bibinfo {author} {\bibfnamefont {H.~E.}\ \bibnamefont {Nistazakis}}, \bibinfo {author} {\bibfnamefont {D.~J.}\ \bibnamefont {Frantzeskakis}},\ and\ \bibinfo {author} {\bibfnamefont {B.~A.}\ \bibnamefont {Malomed}},\ }\bibfield  {title} {\bibinfo {title} {Stabilization of dark solitons in the cubic ginzburg-landau equation},\ }\href {https://doi.org/10.1103/physreve.62.7410} {\bibfield  {journal} {\bibinfo  {journal} {Physical Review E}\ }\textbf {\bibinfo {volume} {62}},\ \bibinfo {pages} {7410–7414} (\bibinfo {year} {2000})}\BibitemShut {NoStop}%
\bibitem [{\citenamefont {Nozaki}\ and\ \citenamefont {Bekki}(1984)}]{Nozaki1984}%
  \BibitemOpen
  \bibfield  {author} {\bibinfo {author} {\bibfnamefont {K.}~\bibnamefont {Nozaki}}\ and\ \bibinfo {author} {\bibfnamefont {N.}~\bibnamefont {Bekki}},\ }\bibfield  {title} {\bibinfo {title} {Exact solutions of the generalized ginzburg-landau equation},\ }\href {https://doi.org/10.1143/jpsj.53.1581} {\bibfield  {journal} {\bibinfo  {journal} {Journal of the Physical Society of Japan}\ }\textbf {\bibinfo {volume} {53}},\ \bibinfo {pages} {1581–1582} (\bibinfo {year} {1984})}\BibitemShut {NoStop}%
\bibitem [{\citenamefont {Malomed}\ and\ \citenamefont {Winful}(1996)}]{Malomed1996}%
  \BibitemOpen
  \bibfield  {author} {\bibinfo {author} {\bibfnamefont {B.~A.}\ \bibnamefont {Malomed}}\ and\ \bibinfo {author} {\bibfnamefont {H.~G.}\ \bibnamefont {Winful}},\ }\bibfield  {title} {\bibinfo {title} {Stable solitons in two-component active systems},\ }\href {https://doi.org/10.1103/physreve.53.5365} {\bibfield  {journal} {\bibinfo  {journal} {Physical Review E}\ }\textbf {\bibinfo {volume} {53}},\ \bibinfo {pages} {5365} (\bibinfo {year} {1996})}\BibitemShut {NoStop}%
\bibitem [{\citenamefont {Kacmoli}\ \emph {et~al.}(2023)\citenamefont {Kacmoli}, \citenamefont {Sivco},\ and\ \citenamefont {Gmachl}}]{Kacmoli2023}%
  \BibitemOpen
  \bibfield  {author} {\bibinfo {author} {\bibfnamefont {S.}~\bibnamefont {Kacmoli}}, \bibinfo {author} {\bibfnamefont {D.~L.}\ \bibnamefont {Sivco}},\ and\ \bibinfo {author} {\bibfnamefont {C.~F.}\ \bibnamefont {Gmachl}},\ }\bibfield  {title} {\bibinfo {title} {Photonic molecule based on coupled ring quantum cascade lasers},\ }\href {https://doi.org/10.1364/optica.497388} {\bibfield  {journal} {\bibinfo  {journal} {Optica}\ }\textbf {\bibinfo {volume} {10}},\ \bibinfo {pages} {1210} (\bibinfo {year} {2023})}\BibitemShut {NoStop}%
\bibitem [{\citenamefont {Kazakov}\ \emph {et~al.}(2024)\citenamefont {Kazakov}, \citenamefont {Letsou}, \citenamefont {Beiser}, \citenamefont {Zhi}, \citenamefont {Opačak}, \citenamefont {Piccardo}, \citenamefont {Schwarz},\ and\ \citenamefont {Capasso}}]{Kazakov2024}%
  \BibitemOpen
  \bibfield  {author} {\bibinfo {author} {\bibfnamefont {D.}~\bibnamefont {Kazakov}}, \bibinfo {author} {\bibfnamefont {T.~P.}\ \bibnamefont {Letsou}}, \bibinfo {author} {\bibfnamefont {M.}~\bibnamefont {Beiser}}, \bibinfo {author} {\bibfnamefont {Y.}~\bibnamefont {Zhi}}, \bibinfo {author} {\bibfnamefont {N.}~\bibnamefont {Opačak}}, \bibinfo {author} {\bibfnamefont {M.}~\bibnamefont {Piccardo}}, \bibinfo {author} {\bibfnamefont {B.}~\bibnamefont {Schwarz}},\ and\ \bibinfo {author} {\bibfnamefont {F.}~\bibnamefont {Capasso}},\ }\bibfield  {title} {\bibinfo {title} {Active mid-infrared ring resonators},\ }\href {https://doi.org/10.1038/s41467-023-44628-7} {\bibfield  {journal} {\bibinfo  {journal} {Nature Communications}\ }\textbf {\bibinfo {volume} {15}} (\bibinfo {year} {2024})}\BibitemShut {NoStop}%
\bibitem [{\citenamefont {Burghoff}\ \emph {et~al.}(2015)\citenamefont {Burghoff}, \citenamefont {Yang}, \citenamefont {Hayton}, \citenamefont {Gao}, \citenamefont {Reno},\ and\ \citenamefont {Hu}}]{Burghoff2015}%
  \BibitemOpen
  \bibfield  {author} {\bibinfo {author} {\bibfnamefont {D.}~\bibnamefont {Burghoff}}, \bibinfo {author} {\bibfnamefont {Y.}~\bibnamefont {Yang}}, \bibinfo {author} {\bibfnamefont {D.~J.}\ \bibnamefont {Hayton}}, \bibinfo {author} {\bibfnamefont {J.-R.}\ \bibnamefont {Gao}}, \bibinfo {author} {\bibfnamefont {J.~L.}\ \bibnamefont {Reno}},\ and\ \bibinfo {author} {\bibfnamefont {Q.}~\bibnamefont {Hu}},\ }\bibfield  {title} {\bibinfo {title} {Evaluating the coherence and time-domain profile of quantum cascade laser frequency combs},\ }\href {https://doi.org/10.1364/oe.23.001190} {\bibfield  {journal} {\bibinfo  {journal} {Optics Express}\ }\textbf {\bibinfo {volume} {23}},\ \bibinfo {pages} {1190} (\bibinfo {year} {2015})}\BibitemShut {NoStop}%
\bibitem [{\citenamefont {Stibenz}\ \emph {et~al.}(2006)\citenamefont {Stibenz}, \citenamefont {Ropers}, \citenamefont {Lienau}, \citenamefont {Warmuth}, \citenamefont {Wyatt}, \citenamefont {Walmsley},\ and\ \citenamefont {Steinmeyer}}]{Stibenz2006}%
  \BibitemOpen
  \bibfield  {author} {\bibinfo {author} {\bibfnamefont {G.}~\bibnamefont {Stibenz}}, \bibinfo {author} {\bibfnamefont {C.}~\bibnamefont {Ropers}}, \bibinfo {author} {\bibfnamefont {C.}~\bibnamefont {Lienau}}, \bibinfo {author} {\bibfnamefont {C.}~\bibnamefont {Warmuth}}, \bibinfo {author} {\bibfnamefont {A.}~\bibnamefont {Wyatt}}, \bibinfo {author} {\bibfnamefont {I.}~\bibnamefont {Walmsley}},\ and\ \bibinfo {author} {\bibfnamefont {G.}~\bibnamefont {Steinmeyer}},\ }\bibfield  {title} {\bibinfo {title} {Advanced methods for the characterization of few-cycle light pulses: a comparison},\ }\href {https://doi.org/10.1007/s00340-006-2190-5} {\bibfield  {journal} {\bibinfo  {journal} {Applied Physics B}\ }\textbf {\bibinfo {volume} {83}},\ \bibinfo {pages} {511–519} (\bibinfo {year} {2006})}\BibitemShut {NoStop}%
\bibitem [{\citenamefont {Han}\ \emph {et~al.}(2020)\citenamefont {Han}, \citenamefont {Ren},\ and\ \citenamefont {Burghoff}}]{han2020sensitivity}%
  \BibitemOpen
  \bibfield  {author} {\bibinfo {author} {\bibfnamefont {Z.}~\bibnamefont {Han}}, \bibinfo {author} {\bibfnamefont {D.}~\bibnamefont {Ren}},\ and\ \bibinfo {author} {\bibfnamefont {D.}~\bibnamefont {Burghoff}},\ }\bibfield  {title} {\bibinfo {title} {Sensitivity of swift spectroscopy},\ }\href {https://doi.org/10.1364/oe.382243} {\bibfield  {journal} {\bibinfo  {journal} {Optics Express}\ }\textbf {\bibinfo {volume} {28}},\ \bibinfo {pages} {6002} (\bibinfo {year} {2020})}\BibitemShut {NoStop}%
\bibitem [{\citenamefont {Busch}\ and\ \citenamefont {Anglin}(2001)}]{busch2001dark}%
  \BibitemOpen
  \bibfield  {author} {\bibinfo {author} {\bibfnamefont {T.}~\bibnamefont {Busch}}\ and\ \bibinfo {author} {\bibfnamefont {J.~R.}\ \bibnamefont {Anglin}},\ }\bibfield  {title} {\bibinfo {title} {Dark-bright solitons in inhomogeneous bose-einstein condensates},\ }\href {https://doi.org/10.1103/physrevlett.87.010401} {\bibfield  {journal} {\bibinfo  {journal} {Physical Review Letters}\ }\textbf {\bibinfo {volume} {87}} (\bibinfo {year} {2001})}\BibitemShut {NoStop}%
\bibitem [{\citenamefont {Hokmabadi}\ \emph {et~al.}(2019)\citenamefont {Hokmabadi}, \citenamefont {Schumer}, \citenamefont {Christodoulides},\ and\ \citenamefont {Khajavikhan}}]{Hokmabadi2019}%
  \BibitemOpen
  \bibfield  {author} {\bibinfo {author} {\bibfnamefont {M.~P.}\ \bibnamefont {Hokmabadi}}, \bibinfo {author} {\bibfnamefont {A.}~\bibnamefont {Schumer}}, \bibinfo {author} {\bibfnamefont {D.~N.}\ \bibnamefont {Christodoulides}},\ and\ \bibinfo {author} {\bibfnamefont {M.}~\bibnamefont {Khajavikhan}},\ }\bibfield  {title} {\bibinfo {title} {Non-hermitian ring laser gyroscopes with enhanced sagnac sensitivity},\ }\href {https://doi.org/10.1038/s41586-019-1780-4} {\bibfield  {journal} {\bibinfo  {journal} {Nature}\ }\textbf {\bibinfo {volume} {576}},\ \bibinfo {pages} {70–74} (\bibinfo {year} {2019})}\BibitemShut {NoStop}%
\bibitem [{\citenamefont {Picqué}\ and\ \citenamefont {H\"{a}nsch}(2019)}]{Picqu2019}%
  \BibitemOpen
  \bibfield  {author} {\bibinfo {author} {\bibfnamefont {N.}~\bibnamefont {Picqué}}\ and\ \bibinfo {author} {\bibfnamefont {T.~W.}\ \bibnamefont {H\"{a}nsch}},\ }\bibfield  {title} {\bibinfo {title} {Frequency comb spectroscopy},\ }\href {https://doi.org/10.1038/s41566-018-0347-5} {\bibfield  {journal} {\bibinfo  {journal} {Nature Photonics}\ }\textbf {\bibinfo {volume} {13}},\ \bibinfo {pages} {146–157} (\bibinfo {year} {2019})}\BibitemShut {NoStop}%
\bibitem [{\citenamefont {Mittal}\ \emph {et~al.}(2021)\citenamefont {Mittal}, \citenamefont {Moille}, \citenamefont {Srinivasan}, \citenamefont {Chembo},\ and\ \citenamefont {Hafezi}}]{mittal2021topological}%
  \BibitemOpen
  \bibfield  {author} {\bibinfo {author} {\bibfnamefont {S.}~\bibnamefont {Mittal}}, \bibinfo {author} {\bibfnamefont {G.}~\bibnamefont {Moille}}, \bibinfo {author} {\bibfnamefont {K.}~\bibnamefont {Srinivasan}}, \bibinfo {author} {\bibfnamefont {Y.~K.}\ \bibnamefont {Chembo}},\ and\ \bibinfo {author} {\bibfnamefont {M.}~\bibnamefont {Hafezi}},\ }\bibfield  {title} {\bibinfo {title} {Topological frequency combs and nested temporal solitons},\ }\href {https://doi.org/10.1038/s41567-021-01302-3} {\bibfield  {journal} {\bibinfo  {journal} {Nature Physics}\ }\textbf {\bibinfo {volume} {17}},\ \bibinfo {pages} {1169–1176} (\bibinfo {year} {2021})}\BibitemShut {NoStop}%
\bibitem [{\citenamefont {Choi}\ \emph {et~al.}(2021)\citenamefont {Choi}, \citenamefont {Hayenga}, \citenamefont {Liu}, \citenamefont {Parto}, \citenamefont {Bahari}, \citenamefont {Christodoulides},\ and\ \citenamefont {Khajavikhan}}]{Choi2021}%
  \BibitemOpen
  \bibfield  {author} {\bibinfo {author} {\bibfnamefont {J.-H.}\ \bibnamefont {Choi}}, \bibinfo {author} {\bibfnamefont {W.~E.}\ \bibnamefont {Hayenga}}, \bibinfo {author} {\bibfnamefont {Y.~G.~N.}\ \bibnamefont {Liu}}, \bibinfo {author} {\bibfnamefont {M.}~\bibnamefont {Parto}}, \bibinfo {author} {\bibfnamefont {B.}~\bibnamefont {Bahari}}, \bibinfo {author} {\bibfnamefont {D.~N.}\ \bibnamefont {Christodoulides}},\ and\ \bibinfo {author} {\bibfnamefont {M.}~\bibnamefont {Khajavikhan}},\ }\bibfield  {title} {\bibinfo {title} {Room temperature electrically pumped topological insulator lasers},\ }\href {https://doi.org/10.1038/s41467-021-23718-4} {\bibfield  {journal} {\bibinfo  {journal} {Nature Communications}\ }\textbf {\bibinfo {volume} {12}} (\bibinfo {year} {2021})}\BibitemShut {NoStop}%
\bibitem [{\citenamefont {Bandres}\ \emph {et~al.}(2018)\citenamefont {Bandres}, \citenamefont {Wittek}, \citenamefont {Harari}, \citenamefont {Parto}, \citenamefont {Ren}, \citenamefont {Segev}, \citenamefont {Christodoulides},\ and\ \citenamefont {Khajavikhan}}]{Bandres2018}%
  \BibitemOpen
  \bibfield  {author} {\bibinfo {author} {\bibfnamefont {M.~A.}\ \bibnamefont {Bandres}}, \bibinfo {author} {\bibfnamefont {S.}~\bibnamefont {Wittek}}, \bibinfo {author} {\bibfnamefont {G.}~\bibnamefont {Harari}}, \bibinfo {author} {\bibfnamefont {M.}~\bibnamefont {Parto}}, \bibinfo {author} {\bibfnamefont {J.}~\bibnamefont {Ren}}, \bibinfo {author} {\bibfnamefont {M.}~\bibnamefont {Segev}}, \bibinfo {author} {\bibfnamefont {D.~N.}\ \bibnamefont {Christodoulides}},\ and\ \bibinfo {author} {\bibfnamefont {M.}~\bibnamefont {Khajavikhan}},\ }\bibfield  {title} {\bibinfo {title} {Topological insulator laser: Experiments},\ }\href {https://doi.org/10.1126/science.aar4005} {\bibfield  {journal} {\bibinfo  {journal} {Science}\ }\textbf {\bibinfo {volume} {359}},\ \bibinfo {pages} {6381} (\bibinfo {year} {2018})}\BibitemShut {NoStop}%
\bibitem [{\citenamefont {Flower}\ \emph {et~al.}(2024)\citenamefont {Flower}, \citenamefont {Jalali~Mehrabad}, \citenamefont {Xu}, \citenamefont {Moille}, \citenamefont {Suarez-Forero}, \citenamefont {\"{O}rsel}, \citenamefont {Bahl}, \citenamefont {Chembo}, \citenamefont {Srinivasan}, \citenamefont {Mittal},\ and\ \citenamefont {Hafezi}}]{Flower2024}%
  \BibitemOpen
  \bibfield  {author} {\bibinfo {author} {\bibfnamefont {C.~J.}\ \bibnamefont {Flower}}, \bibinfo {author} {\bibfnamefont {M.}~\bibnamefont {Jalali~Mehrabad}}, \bibinfo {author} {\bibfnamefont {L.}~\bibnamefont {Xu}}, \bibinfo {author} {\bibfnamefont {G.}~\bibnamefont {Moille}}, \bibinfo {author} {\bibfnamefont {D.~G.}\ \bibnamefont {Suarez-Forero}}, \bibinfo {author} {\bibfnamefont {O.}~\bibnamefont {\"{O}rsel}}, \bibinfo {author} {\bibfnamefont {G.}~\bibnamefont {Bahl}}, \bibinfo {author} {\bibfnamefont {Y.}~\bibnamefont {Chembo}}, \bibinfo {author} {\bibfnamefont {K.}~\bibnamefont {Srinivasan}}, \bibinfo {author} {\bibfnamefont {S.}~\bibnamefont {Mittal}},\ and\ \bibinfo {author} {\bibfnamefont {M.}~\bibnamefont {Hafezi}},\ }\bibfield  {title} {\bibinfo {title} {Observation of topological frequency combs},\ }\href {https://doi.org/10.1126/science.ado0053} {\bibfield  {journal} {\bibinfo  {journal} {Science}\ }\textbf {\bibinfo {volume} {384}},\ \bibinfo {pages} {1356–1361} (\bibinfo {year}
  {2024})}\BibitemShut {NoStop}%
\end{thebibliography}%

\newpage

\begin{acronym}

\acro{mir}[mid-IR]{mid-infrared}
\acro{qwip}[QWIP]{quantum well infrared photodetector}
\acro{lo}[LO]{local oscillator}
\acro{ftir}[FTIR]{Fourier transform infrared}
\acro{swifts}[SWIFTS]{Shifted-Wave Interference Fourier Transform
Spectroscopy}
\acro{fsr}[FSR]{free spectral range}
\acro{rf}[RF]{radio frequency}
\acro{nb}[NB]{Nozaki-Bekki}
\acro{cw}[CW]{continuous-wave}
\acro{qcl}[QCL]{quantum cascade laser}
\acro{qcls}[QCLs]{quantum cascade lasers}
\acro{ht}[HT]{heater}
\acro{fp}[FP]{Fabry-P\'{e}rot}
\acro{shb}[SHB]{spatial hole burning}
\acro{cgle}[CGLE]{complex Ginzburg-Landau equation}
\acro{gvd}[GVD]{group velocity dispersion}
\acro{lef}[LEF]{linewidth enhancement factor}
\acro{rt}[RT]{racetrack}
\acro{rts}[RTs]{racetracks}
\acro{wg}[WG]{waveguide}
\acro{gdd}[GDD]{group delay dispersion}
\acro{cmt}[CMT]{coupled mode theory}
\acro{ep}[EPs]{exceptional points}
\acro{mct}[MCT]{Mercury Cadmium Telluride}
\acro{dc}[DC]{direct current}

\end{acronym}

\end{document}